\documentclass{mem}
\usepackage{graphicx}
\usepackage{times}
\usepackage{balance}
\usepackage{natbib}
\usepackage{txfonts}
\usepackage{subfigure}
\usepackage[a4paper]{hyperref}

\def\eg{{\it e.g.,~}}
\def\ie{{\it i.e.,~}}
\def\ltsim{\; \raise0.3ex\hbox{$<$\kern-0.75em \raise-1.1ex\hbox{$\sim$}}\; }
\def\gtsim{\; \raise0.3ex\hbox{$>$\kern-0.75em \raise-1.1ex\hbox{$\sim$}}\; }

\begin{document}

\title{Radio Halos, cluster mergers and the role of future LOFAR observations}
\subtitle{}

\author{R. Cassano}

\offprints{rcassano@ira.inaf.it}
\institute{INAF, Istituto di Radioastronomia, via P. Gobetti 101, 4014, 
Bologna (Italy) \\
Dipartimento di Astronomia,Universita' di Bologna, via Ranzani 1, 
I-40127 Bologna, Italy\\}

\abstract{A radio bimodality is observed in galaxy clusters: 
a fraction of clusters host giant radio halos while the majority 
of clusters do not show evidence of diffuse cluster-scale radio emission. 
Present data clearly suggest that the radio bimodality has a 
correspondence in terms of dynamical state of the hosting clusters. 
I will report on these evidences in some details and discuss the role of 
cluster mergers in the generation of giant radio halos and their evolution. 
Finally I will report on expectations on the statistical
properties of radio halos assuming that the
emitting electrons are re-accelerated by merger-turbulence, 
and discuss the role of incoming LOFAR surveys.}

\titlerunning{Radio halos and cluster mergers}
\authorrunning{R. Cassano}

\maketitle

\section{Introduction}\label{sec:intro}

Radio and X-ray observations of galaxy clusters prove that thermal 
and non-thermal components coexist in the intra-cluster medium (ICM). 
While X-ray observations reveal thermal emission from diffuse hot gas, 
radio observations of an increasing number of massive galaxy clusters 
unveil the presence of ultra-relativistic particles and magnetic fields 
through the detection of diffuse, giant Mpc-scale synchrotron 
{\it radio halos} (RH) and {\it radio relics} 
\citep[\eg][]{ferrari08, cassano09}. 
RHs are the most spectacular evidence of non-thermal components in the ICM. 
They are giant radio sources located in the cluster central regions, 
with spatial extent similar to that of the hot ICM and steep radio 
spectra, $\alpha\simeq 1.2-1.5$  \citep[\eg][this conference]{venturi11}.

There are well known correlations between the synchrotron monochromatic 
radio luminosity of RH ($P_{1.4\,\mathrm{GHz}}$) and 
the host cluster X-ray luminosity ($L_X$), mass and 
temperature \citep[\eg][] {liang00, feretti03, cassano06, brunetti09}.
The most powerful RH are found in the most X-ray luminous, massive and 
hot clusters. These correlations suggest a close link between the 
non-thermal and the thermal/gravitational cluster physics.  

Another important fact is that RHs are 
found in clusters that show recent/ongoing merging activity: 
significant substructure and distortion in the X-ray 
images \citep[\eg][] {sch01}, complex gas temperature 
distributions \citep[\eg][] {govoni04, bourdin11}, 
shocks and cold fronts  \citep[\eg][] {markevitch01,markevitch10},
absence of strong cooling-flow 
\citep[\eg][this conference] {feretti03, rossetti11}, and 
optical substructures \citep[\eg][]{boschin06}.

In a seminal paper \cite{buote01} provided the first quantitative 
comparison of the dynamical states of clusters with RH and the properties
of RHs and discovered a correlation between the RH luminosity at 
1.4 GHz and the magnitude of the dipole power ratio 
$P_1/P_0$. This implies that the more powerful RHs are hosted in clusters
that experience the largest departures from virialization.

The RH-merger connection and the thermal--non-thermal correlations suggest
that the gravitational process of cluster formation may provide the energy 
to generate the non-thermal components in clusters through the acceleration 
of high-energy particles via shocks and turbulence (\eg Sarazin 2004,
Brunetti 2011, this conference). 
Cluster-cluster mergers are among the most energetic events in the present 
Universe: two clusters with total masses $M_1$ and $M_2$ dissipate a 
gravitational energy (in {\rm erg}) : 

\begin{equation}
E_{\rm g} \approx 
10^{64}
\bigg(\frac{M_{1}}{10^{15}\,\mathrm{M_{\odot}}}\bigg)
\bigg(\frac{M_{2}}{10^{15}\,\mathrm{M_{\odot}}}\bigg)
\bigg(\frac{d_o}{6 \mathrm{Mpc}}\bigg)^{-1} 
\end{equation}

\noindent $d_o$ is the turnaround distance.
The theoretical goal is 
to understand how a fraction of this large energy budget is channeled 
into the acceleration of high energy particles and amplification 
of cluster magnetic field (\eg Brunetti 2011, this conference). 

The recent discovery of RHs with very steep spectrum provides additional
support to the scenario where RHs are generated due to re-acceleration
of relativistic particles by merger-driven turbulence 
(\eg Brunetti et al 2008, Brunetti 2011 this conference).
Future low-frequency radio telescopes (such as LOFAR and LWA) have the 
potential to test this scenario and to further explore the connection 
between RH and the process of cluster formation. 
Here I will discuss the most recent evidences in favor of the connection 
between RHs and clusters mergers and the expectations of the turbulent 
re-acceleration scenario to test trough future low frequency observations.

\begin{figure*}\label{LrLx}
\begin{center}
\includegraphics[width=0.8\textwidth]{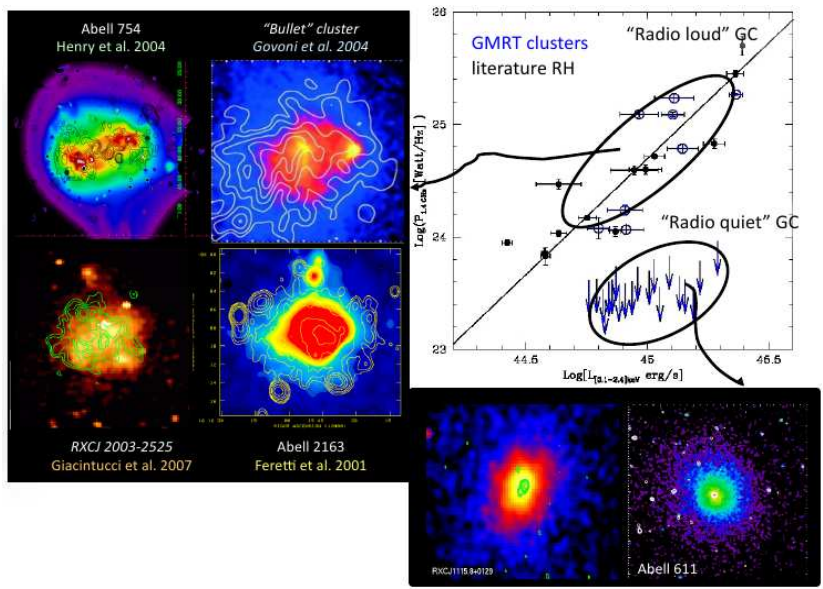}
\caption{\footnotesize
Distribution of clusters in the plane radio ($P_{1.4}$) -- X-ray luminosity 
($L_{[0.1-2.4] keV}$) for clusters of the GMRT RH Survey (blue symbols) 
and for
RH from the literature (black dots); adapted from Brunetti et al.(2009).}
\end{center}
\end{figure*}

\section{The GMRT RH Survey and the radio bi-modality of clusters}

Recently, deep radio observations of a complete sample of galaxy clusters 
have been carried out as part of the Giant Metrewave Radio Telescope 
({\it GMRT}) RH Survey \citep{venturi07,venturi08}. 
These observations confirmed that diffuse cluster-scale radio emission 
is not ubiquitous in clusters: only ~30\% of the X-ray luminous 
($L_X(0.1-2.4\,\mathrm{keV})\geq5\times10^{44}$ erg/s) clusters host a RH. 
Most importantly, these observations allow to separate RH clusters from 
clusters without RH, showing a bimodal distribution of clusters in 
the $P_{1.4\,{\mathrm GHz}}$ -- $L_X$ diagram \citep{brunetti07}: 
RHs trace the well known correlation between $P_{1.4\,{\mathrm GHz}}$ 
and $L_X$, while the upper limits to the radio luminosity of clusters 
with no-RH lie about one order of magnitude below 
that correlation (Fig.~\ref{LrLx}). 
Why clusters with the same thermal properties (and at the same cosmological 
epoch) have different non-thermal properties ? 
One possibility, which was first suggested by Venturi et al. (2008) based 
on information from the literature available for a fraction of
the clusters of the GMRT RH Survey, 
is that the behavior of clusters in the 
$P_{1.4\,{\mathrm GHz}}-L_X$ diagram is connected with their dynamical state;
this is supported also by a simple visual inspection of the X-ray images 
of those clusters (Fig.~\ref{LrLx}).

\begin{figure}
\label{substructures}
\begin{center}
\includegraphics[width=0.55\textwidth]{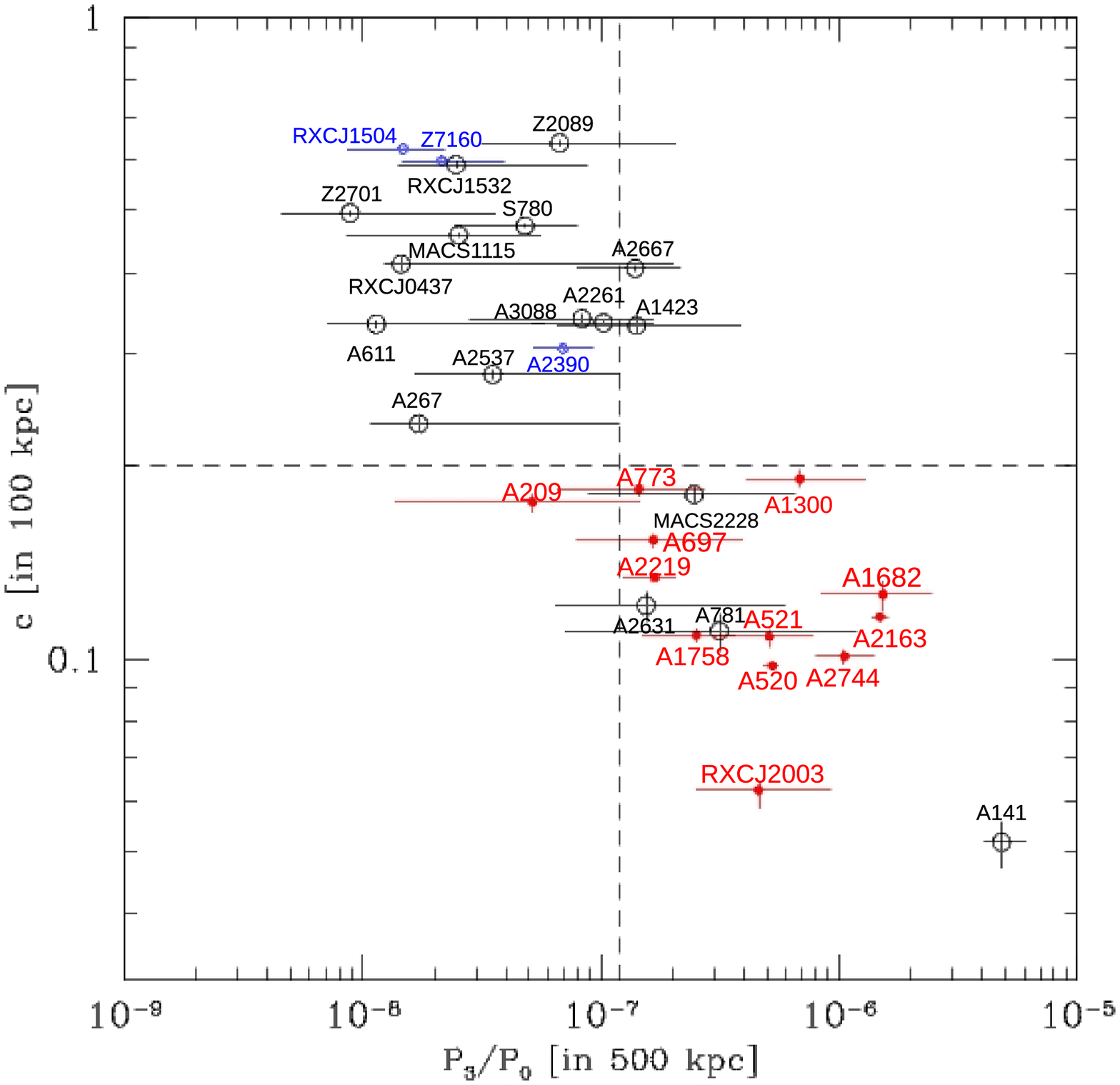}
\includegraphics[width=0.55\textwidth]{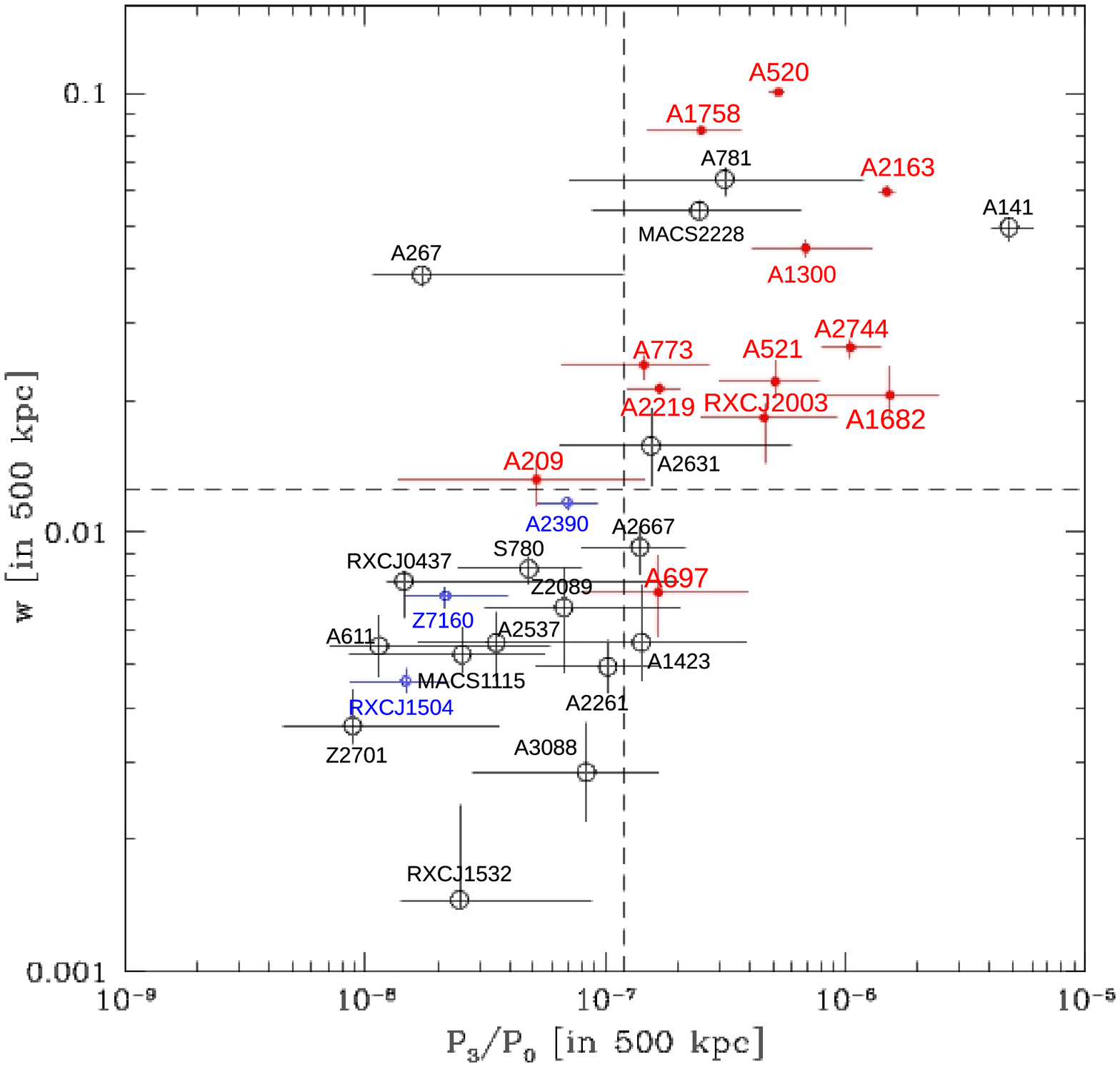}
\caption{\footnotesize
{\bf Upper Panel}: concentration parameter c vs. power ratio P3/P0; 
{\bf Lower Panel}: 
centroid shift w vs. P3/P0. 
Symbols are: RH (red filled dots), non-RH (black
open dots), mini-halos (blue open dots). 
Vertical and horizontal dashed lines mark : 
$c = 0.2$, $w = 0.012$ and $P_3/P_0=1.2\times10^{-7}$.
}
\end{center}
\end{figure}

\section{Dynamical state of GMRT clusters}
\label{sec:dyn_cluster}

In a more recent work \cite{cassano10a} using {\it Chandra} archive X-ray 
data of a sub-sample of clusters belonging to the GMRT RH 
Survey\footnote{Those with archival data with at least 2000 ACIS-S 
or ACIS-I counts in the 0.5-2 keV band inside an aperture 
of 500 kpc \citep[see][]{cassano10a}.} provided a more quantitative measure 
of the degree of the cluster disturbance using three different methods: 
power ratios \citep[\eg][]{buote95, jeltema05}, the emission centroid 
shift \citep[\eg][]{mohr93,poole06}, and the surface brightness concentration 
parameter \citep[\eg][]{santos08}. 
The power ratio method is a multipole decomposition of the gravitational 
potential of the two-dimensional projected mass distribution inside a given 
aperture $R_{ap}$. Following \cite{buote95} they are usually defined as 
$P_m/P_0$, where $P_m$ represents the square of the mth multipole of the 
two-dimensional potential \citep[see eqs. 1-4 in][]{cassano10a}.  
Large departures from a virialized state are then indicated by large power 
ratios. 

The centroid shift, $w$, is defined as the standard deviation of the 
projected separation between the peak of the X-ray emission and the centroid, 
derived in increasing circular apertures and expressed in units  
of $R_{ap}=500$ kpc. 
The centroid shift, $w$, 
is a measure of the skewness of the photon distribution of a cluster, 
thus larger values of $w$ indicate clusters with a more 
asymmetric/irregular distribution of the X-ray emission.

The concentration parameter, $c$, defined as the ratio of the peak over 
the ambient surface brightness, $S$, 
$c=\frac{S(r<100\,\mathrm{kpc})}{S(<500\,\mathrm{kpc})}$, has been used 
in literature for identification of cool core clusters \citep{santos08}. 
We used $c$ to separate galaxy clusters with a compact core (higher 
values of $c$, core not disrupted from recent merger events)
from clusters with a broad distribution of the gas in the core
(lower values of $c$, core disturbed from a recent merger episode).  

\begin{figure}
\label{spettri}
\resizebox{\hsize}{!}{\includegraphics[clip=true]{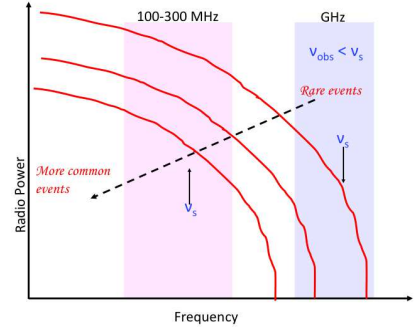}}
\caption{\footnotesize
Schematic representation of the synchrotron spectra of RHs with different 
values of $\nu_s$ (see text for details). Those with higher values 
of $\nu_s$ are visible up to GHz 
frequencies, while those with lower values would be observable only at 
lower frequencies.}
\end{figure}

\subsection{Results}
\label{sec:results}
In Fig.\ref{substructures} we report the distribution of the 32 clusters 
in the ($c,P_3/P_0$) plane (upper panel) and in the ($w,P_3/P_0$) plane 
(lower panel). We found that RH clusters (red filled dots) can be well 
separated from clusters without RH (black open dots) and clusters with 
mini-halos\footnote{Radio mini-halos are diffuse synchrotron emission 
on smaller scales (\eg 200-500 kpc) extending around powerful radio galaxies 
at the center of some cool core clusters (\eg Venturi 2011, this conference).}
(blue open dots). 
Clusters with RHs are found only in the region of low values of 
$c$ ($c\ltsim0.2$), and high values of $P_3/P_0$ ($P_3/P_0 \gtsim 
1.2\times 10^{-7}$) and $w$ ($w\gtsim0.012$). 

Both diagrams provide strong evidence that RHs form in dynamical disturbed 
clusters, while clusters with no evidence of Mpc-scale synchrotron emission 
are more relaxed systems. 
We also tested quantitatively this result by running Monte Carlo simulations 
(see Cassano et al. 2010a for details) and proved that the observed 
distribution differs from a random one (\ie independent of cluster dynamics) 
at more than $4\sigma$. 
This proves that our result is statistically significant and shows, 
for the first time, that the separation between RHs and non-RH clusters 
(the observed radio bi-modality of clusters) has a corresponding separation 
in terms of dynamical properties of the host clusters.
We note that there are 4 outliers in Fig.\ref{substructures}: Abell 781, 
MACS 2228, Abell 141 and Abell 2631, \ie clusters that are dynamically 
disturbed but that do not host a RH. 
\footnote{For one of them, Abell 781, more recently we verified that its 
0.2-2.4 keV luminosity, $L_X \sim 4-5\times 10^{44}$ erg/s, 
\citep[\eg][S. Ettori, private communication, based on a shallow ROSAT HRI
exposure]{boe00,maughan08} is substantially smaller than that used in the GMRT
sample (taken from the \cite{eb98} catalogue, $L_X \simeq 1.13\times10^{45}$
erg/s) and is smaller than the threshold value ($L_X\gtsim5\times10^{44}$
erg/s) used to select clusters for the GMRT RH Survey.}
These clusters deserve further investigation (see also next Sect.).

\section{Turbulent re-acceleration scenario and low frequency observations}

A promising scenario proposed to explain the origin of the synchrotron 
emitting electrons in RHs assumes that electrons are re-accelerated due to 
the interaction with MHD turbulence injected in the ICM during 
cluster mergers ({\it turbulent re-acceleration} model, 
\citep[\eg][]{brunetti01,petrosian01}. 

\noindent
This scenario naturally explain the observed bi-modality of 
clusters \citep[\eg][]{brunetti09} and the observed connection between 
RHs and cluster mergers.

Stochastic particle acceleration by MHD turbulence is rather inefficient in 
the ICM, consequently electrons can be accelerated only up to energies 
of $m_e c^2 \gamma_{max} \leq$ several GeV. 
This entails a high-frequency cut-off in the synchrotron spectra of RHs, 
which marks the most important expectation of this scenario.
The presence of this cut-off implies that the observed fraction of clusters 
with RHs depends on the observing frequency, this 
can be immediately understood from Fig.~\ref{spettri}. 
The steepening of the spectrum makes it difficult to detect RHs 
at observing frequencies larger than the frequency, $\nu_s$, where 
the steepening becomes severe. 
The frequency $\nu_s$ depends on the acceleration efficiency in the ICM, 
which in turns depends on the flux of MHD turbulence dissipated in 
relativistic electrons \citep[\eg][]{cassano06,cassano10b}. 
Larger values of $\nu_s$ are expected in more massive clusters and 
in connection with major merger events. 
As a consequence, according to this model, present radio surveys at 
$\sim$ GHz frequencies can reveal only those RHs generated during the 
most energetic merger events and characterized by relatively flat 
spectra ($\alpha \sim 1.1-1.5$) (see Fig.\ref{spettri}). 
These sources should represent the tip of the iceberg of the whole population 
of RHs, since the bulk of cluster formation in the Universe occurs 
trough less energetic mergers. 
Low frequency observations with next generation of radio telescopes 
(LOFAR, LWA) are thus expected to unveil the bulk of RHs, including a 
population of RH which will be observable preferentially at low radio 
frequencies ($\nu \leq 200-300$ MHz). 
These RHs, generated during less energetic but more common merger 
events, should have extremely steep radio spectra ($\alpha \gtsim 1.5-1.9$)
when observed at higher frequencies;
we defined these sources Ultra Steep Spectrum RH (USSRH). 
Possible prototypes of these RHs are those found in 
Abell 521 \citep[$\alpha\sim 2$][]{brunetti08} and 
in Abell 697 \citep[$\alpha\sim 1.7$][]{macario10}.

In the framework of the {\it turbulent re-acceleration}
scenario, the existence of merging clusters with no Mpc-scale radio 
emission (\eg in Fig.~\ref{substructures}) is not surprising 
for two main reasons. 
First, the expected lifetime of RHs ($\sim$ Gyr) can be smaller than the 
typical time-scale of a merger, during which the cluster would appear 
disturbed, readily implying that not all disturbed systems should host 
RHs \citep[\eg][]{brunetti09}. 

\noindent
Second, and most important, 
a fraction of clusters should host USSRH that are difficult to detect 
trough observations at high frequencies.
USSRH are mainly expected in disturbed clusters with masses 
$M_v \ltsim 10^{15}\,M_{\odot}$ (in the local Universe), or in merging
(massive) clusters at higher redshift, 
$z\gtsim 0.4-0.5$ \citep[][]{cassano10b}. 
In line with this scenario, 3 out of the 4 outliers have X-ray luminosity 
close to the lower boundary used to select the GMRT sample 
($L_X = 5\times 10^{44}$ erg/sec), and the other is the cluster 
with the highest redshift in the GMRT sample ($z \simeq 0.42$). 
Interestingly, a deep GMRT follow-up at 325 MHz of one of 
the outliers in Fig.~\ref{substructures}, Abell 781,  
has revealed the presence of a possible USSRH (Giacintucci 2011, 
this conference; Venturi et al. 2011, submitted), 
which need to be confirmed by future deeper low-frequency observations.

USSRH are expected to be less powerful than RHs with 
flatter spectra \citep[see][]{cassano10} and thus very sensitive 
low-frequency observations are necessary to catch them. 
The ideal instrument to search these RHs is LOFAR (LOw Frequency ARray) that 
is already operating in commissioning phase \citep[\eg][]{roettgering10}. 
To derive quantitatively the statistical properties of RHs, 
we used Monte Carlo procedures \citep[\eg][]{cassano05} that follow the 
process of cluster formation, the injection and dissipation of turbulence 
during cluster-cluster mergers and the ensuing acceleration of relativistic 
particles in the ICM.
The expectations based on these procedures were found consistent 
with present observational constraints \citep[\eg][]{cassano08}. 
Thus using the same procedures we derived expectations for the 
planned LOFAR surveys.
The {\it Tier 1} ``Large Area Survey'' at 120 MHz 
\citep[see][]{roettgering10} is expected to greatly increase the
number of known giant RHs with the possibility to 
detect about 350 RHs up to redshift $z \approx 0.8$ with about half of
these RHs having very steep radio spectra
\citep[$\alpha\gtsim1.9$,][]{cassano10b}.
Consequently future LOFAR surveys will allow a powerful test of the 
merger-driven turbulence re-acceleration scenario for the origin of RHs.

\section{Conclusions}
\label{sec:future}

We discussed the most recent evidences that demonstrate a connection between 
the generation of Mpc-scale radio emission in clusters, in the form of 
giant RHs, and the merging activity in clusters. 

\noindent
A step forward in this direction comes from the discovery that 
the radio bi-modality of clusters
has a correspondence in terms of dynamical state of the clusters : 
clusters with RHs are found to be dynamically disturbed, while clusters 
without RHs are dynamically relaxed. 
This has been proved by applying three different methods to characterize 
cluster substructures to the X-ray Chandra images of GMRT 
clusters \citep{cassano10a}. 

\noindent
The correlation between the synchrotron radio luminosity of RHs and the
cluster X-ray luminosity (mass and temperature) combined with 
the connection between RHs and cluster mergers 
suggest that there are at least two main ingredients in the generation
of RHs : the cluster dynamical status and cluster mass.
These observational facts are naturally understood in the framework of one 
of the proposed pictures put forward to explain the origin of giant RHs, 
the merger-induced turbulence re-acceleration 
scenario \citep{brunetti01,petrosian01}. 
This scenario has unique expectations for the statistical properties of RHs 
that could be tested by future radio surveys at low frequencies.

\noindent
In particular, the shape of the spectrum of RHs is connected with the
energy dissipated during cluster mergers and thus ultimately with the
mass of the hosting clusters and with the mass ratio (impact parameter
etc) of merger events.
A large fraction of RHs, those associated with less massive merging
systems and those at higher redshift, should have ultra-steep spectra
and glow up preferentially in deep surveys at low radio frequencies
\citep{cassano10b}.
LOFAR is thus expected to perform a powerful test of this scenario.

\begin{acknowledgements}
This work is partially supported by INAF under grants PRIN-INAF2008 and
PRIN-INAF2009 and by ASI-INAF under grant I/088/06/0.
I acknowledge my main collaborators S. Ettori, G. Brunetti, S. Giacintucci,
T. Venturi, M. Br\"uggen  and H. R\"ottgering.
\end{acknowledgements}

\bibliographystyle{aa}

\end{document}